\documentclass[conference]{IEEEtran}
\IEEEoverridecommandlockouts
\usepackage{cite}
\usepackage{amsmath,amssymb,amsfonts}
\usepackage{amsthm}

\usepackage{algorithmic}
\usepackage{graphicx}
\usepackage{textcomp}
\usepackage{xcolor}
\def\BibTeX{{\rm B\kern-.05em{\sc i\kern-.025em b}\kern-.08em
    T\kern-.1667em\lower.7ex\hbox{E}\kern-.125emX}}
    
\usepackage{tikz}
\usepackage{textcomp}
\usepackage{hyperref}
\usepackage{lipsum}

\begin{document}

\makeatletter
\newcommand{\linebreakand}{%
  \end{@IEEEauthorhalign}
  \hfill\mbox{}\par
  \mbox{}\hfill\begin{@IEEEauthorhalign}
}
\makeatother

\title{Towards a Cognitive Compute Continuum: An Architecture for Ad-Hoc Self-Managed Swarms}

\author{
\IEEEauthorblockN{Ana Juan Ferrer\IEEEauthorrefmark{1}, Soeren Becker\IEEEauthorrefmark{3}, Florian Schmidt\IEEEauthorrefmark{3}, Lauritz Thamsen\IEEEauthorrefmark{3}, and Odej Kao\IEEEauthorrefmark{3}}

\IEEEauthorblockA{
\IEEEauthorrefmark{1}
ajuanf@uoc.edu, Universitat Oberta de Catalunya, Atos, Spain\\
}
\IEEEauthorblockA{
\IEEEauthorrefmark{3}
\{soeren.becker, florian.schmidt, lauritz.thamsen, odej.kao\}@tu-berlin.de, Technische Universität Berlin, Germany\\
}
}

\maketitle

\begin{abstract}
In this paper we introduce our vision of a Cognitive Computing Continuum to address the changing IT service provisioning towards a distributed, opportunistic, self-managed collaboration between heterogeneous devices outside the traditional data centre boundaries. The focal point of this continuum are cognitive devices, which have to make decisions autonomously using their on-board computation and storage capacity based on information sensed from their environment. Such devices are moving and cannot rely on fixed infrastructure elements, but instead realise on-the-fly networking and thus frequently join and leave temporal swarms. All this creates novel demands for the underlying architecture and resource management, which must bridge the gap from edge to cloud environments, while keeping the QoS parameters within required boundaries. The paper presents an initial architecture and a resource management framework for the implementation of this type of IT service provisioning.
\end{abstract}

\section{Introduction}
The ongoing evolution of IT provisioning disrupts the traditional way of computation and storage, where services are bound to massive data centres, secured by firewalls, maintained professionally, and focused on keeping the application logic and intelligence within the data centre, while the outside world is limited to the role of clients. The introduction of smartphones and mobile networks removed many of the spatial constraints, but the usage still relies on conventional request/reply protocols. However, the need for low-latencies exposed the necessity to move the processing closer to the origin of the data and brought forward manifold concepts such as edge and fog computing. This enabled a variety of novel autonomous applications and methods involving sensors, actuators, robots, drones, cars, routers, and servers and leading to scalability demands, which exceed all previous assumptions. A significant conceptual change results from the elimination of the fixed infrastructures – a gateway, a transmission tower, an access point – and towards an ad-hoc and meshed networking on a high pace and frequency. In contrast to previous distributed computing concepts, those devices do not mesh solely for the sake of connectivity but aim at exchanging acquired status information on the computing environment and at making decisions that are critical for the correct functioning of the entire system. IoT is one of the major developments leading to more complex devices, for example self-driving cars or drones. Such devices have significant computing and storage power on-board, can acquire information, and implement reactions. They therefore represent a novel type of sense-and-act devices that we name \textit{cognitive resources}. They are not constrained to any data centre boundary, but located somewhere in an open space. Consequently, they are vulnerable to malfunctions due to environmental influences, commodity maintenance, as well as attacks and destruction. These properties and the need for fast, ad-hoc, and binding decisions requires a novel type of resource management and an architecture for cognitive devices that is able to implement the required on-demand, opportunistic, ad-hoc management approach.

In this paper we introduce our vision for Cognitive Computing Continuums (CCCs) alongside a general architecture for implementing the necessary dynamic resource management. We envision CCCs as a \textit{distributed, opportunistic, collaborative, heterogeneous, self-managed, sensing, and learning environment, bridging the Edge and Clouds}. Thus, 
CCCs will enable service provisioning to shift from the existing predefined and fixed orientation to a pure on-demand, opportunistic, and ad-hoc approach. In this approach, executing services will rely on dynamic, diverse, and distributed cognitive resources. These cognitive resources can belong to different domains and may have possibly conflicting interests. 
This approach aims at taking full advantage of all available heterogeneous resources to provide a service-based environment that allows for collecting and utilising these massive, diverse, and highly distributed cognitive resources. 
We name these distributed service networks \textit{swarms} and envision them as opportunistic, created on-the-fly in order to respond to specific current needs. Coordination of \textit{swarms} occurs in an open and distributed self-management runtime model. We expect this self-management model to take advantage of learning techniques for its control and administration --- i.e. utilising reservation, negotiation, adaptive selection, conflict resolution --- and to apply techniques for anticipating the volatility and uncertainty introduced by real-world availability of dynamic cognitive resources, thereby enabling an efficient and reliable service provisioning.
CCC scenarios will need to address self-management and self-healing at scale and across heterogeneous cognitive resources and clouds types, but also need to consider business models and incentives that regulate the participation of different entities in \textit{swarms}. 

Translating those requirements into a general architecture results into the guiding idea of an extended, context-aware overlay P2P-based network controlled by a middleware, which bootstraps the building of \textit{swarms} and coordinates the activities within swarms to provide the required Quality of Service (QoS) despite expected node fluctuations and service disruptions.  

The remaining of the paper is organised as follows. A survey and an analysis of current cloud-based infrastructures in Section \ref{stateCC} outlines the necessity for the introduction of Cognitive Cloud Continuums to react to changing demands. Section \ref{principlesCCC} presents the central CCC characteristics and introduces their three key properties \textit{awareness, autonomous, actionable}. Finally, Section \ref{CCRealisation} provides an initial idea and an approach for the realisation of the vision of CCCs. An overview of the related work and a conclusion complete the paper.

\section{Current State of Cloud Environments}\label{stateCC}
The idea of CCCs aims to extend the traditional concepts of cloud environments. The understanding of present concepts in Cloud Computing is therefore needed for overcoming the limitations by introducing CCCs. We therefore describe the current state of Clouds in more detail to show the evolution of Cloud concepts with respect to past and current limitations as well as requirements for future Clouds. Figure \ref{fig:my_label} illustrates the full development of the different concepts of Clouds, while including also our idea of a Cognitive Cloud.
\begin{figure}
    \centering
    \includegraphics[width=0.5\textwidth]{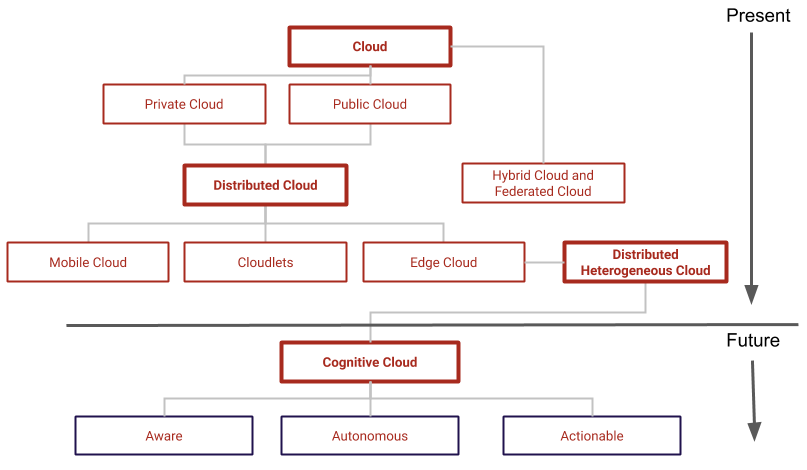}
    \caption{Present and future state of Cloud environments.}
    \label{fig:my_label}
\end{figure}

\subsubsection{Cloud Computing (Traditional Cloud Computing / Multi-Region Cloud Computing)}
Cloud Computing provides the illusion of infinite computing resources which are available on-demand.
First mentioned around 2006 - mainly as a marketing term in various contexts \cite{vaquerobreak2009} - it has enabled a radical acceleration of the commodification of computing and provided the means for new business models and services. 
In Cloud computing scenarios, consumers are enabled to automatically provision computing capabilities without requiring human intervention of the service provider. They can rapidly scale up and down their resources and the usage is transparently monitored for both the provider and consumer of the resource. Large capital outlays in hardware are no longer required which allows companies to significantly decrease their capital expenditures \cite{armbrust_view_2010, zhang2010cloud}.

The main service models of Cloud Computing are outlined as Infrastructure as as Service (IaaS), Platform as a Service (PaaS) and Software as a Service (SaaS): IaaS
refers to the provisioning of essential computing resources such as processing power, (virtual) networks or storage whereas PaaS offers platform-based services, for instance 
container orchestration. The on-demand deployment of specific software in the cloud infrastructure is called SaaS \cite{mell2011nist}.

The NIST definition also included different deployment models of Cloud Computing, namely Private Cloud, Public Cloud and Hybrid Cloud.The latter are often incorporated to complement the on-premise infrastructure of organizations (private cloud) with further computing resources at public clouds\cite{sotomayor2009virtual}.

Nowadays, public cloud providers like Amazon, Google, or Microsoft operate data centres around the world to offer cloud services within a latency-defined perimeter for their users (Multi-region Cloud): Customers can choose the regions in which they want to deploy their services to ensure fast response times and deliver cross-region resiliency.

\subsubsection{Distributed Cloud } %

Distributed Clouds further increase the distribution of public cloud services to different
physical locations and represent a larger set of diverse architectures.
As described by Gartner \cite{Gartner}, the key responsibility for managing physically distributed public clouds services originates at the public cloud provider. This definition holds for Multi-region clouds (cross-regional clouds), which are managed through a centralized virtualization layer.

Multiple clouds in different regions meet demands towards latency preserving SLAs up to political aware demands due to physical locality of the cloud to their consumers.
Technically, Multi-clouds might contain multiple management units/interfaces. Such separation by management units allow building higher levels of management, which can appear as federated or hybrid clouds. The single cloud entities are still managed in a centralized manner and provide mostly homogeneous resources although Gartner's definition focuses on cloud services and does not demand homogeneous cloud resources.

The expected growth of connected things and devices require decentralization of cloud computing in order to avoid unnecessary latencies by distributing computing to the edges of the network.
Compared to a Multi-region cloud computing, distributed cloud computing aims to relax this requirement and provides cloud services additionally on heterogeneous devices.
In the last decades, mainly three different concepts arose to cope with the character of decentralization: Mobile clouds, Cloudlets, Edge/Fog Clouds.

\begin{figure}
    \centering
    \includegraphics[width=0.5\textwidth]{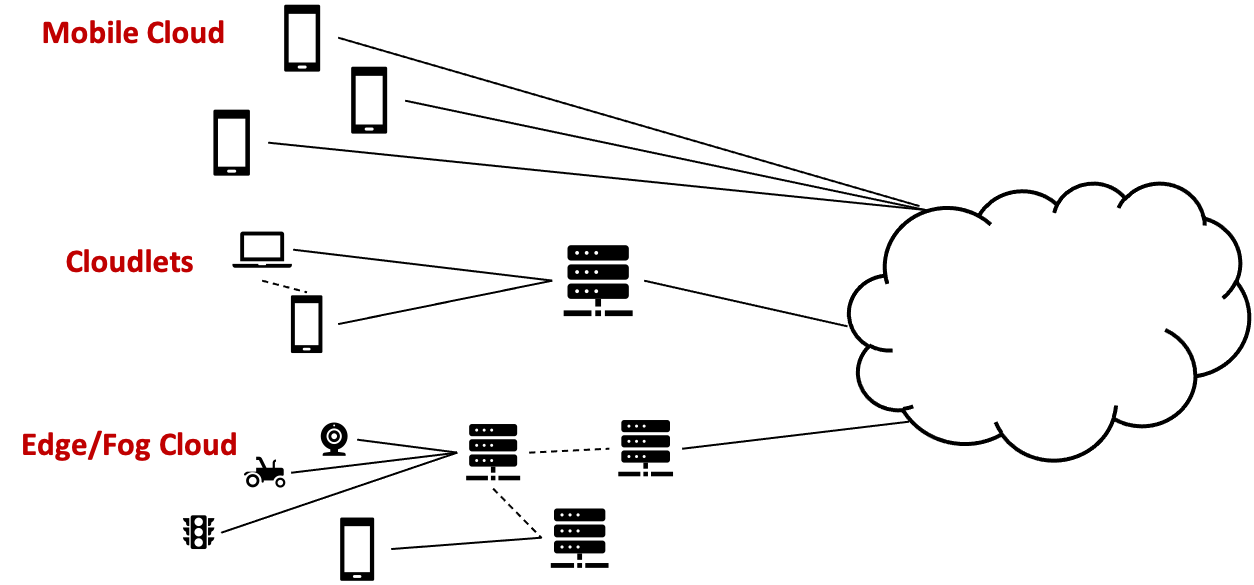}
    \caption{Conceptualizational overview of the three architectures: Mobile cloud, Cloudlets, and Edge/Fog cloud.}
    \label{fig:distributedCloud}
\end{figure}

\paragraph{Mobile cloud}
Mobile clouds (MC) were coined by the emerging smartphones capable to access external compute resources through the internet. As depicted in Figure \ref{fig:distributedCloud}, mobile devices function as thin clients and communicating with additional cloud backend services. As mobile devices does not provide large compute resources, computational extensive applications can be offloaded to the cloud.
In the same notation and as defined by Kovachev et al. \cite{kovachev2011mobile} MC computing describes the augmentation of mobile device capabilities via ubiquitous wireless access to cloud services.

In a nutshell, Mobile cloud computing refers to an
infrastructure where both the data storage and data processing are offloaded form the mobile device. Mobile cloud
applications move the computing power and data storage from mobile phones into the cloud, bringing applications and MC to not just smartphone users but a much broader range of mobile subscribers.

\paragraph{Cloudlets}

In context of mobile clouds, the offloading of compute resources to the cloud bears challenges for real-time applications as cloud data centers are typically located far away from the mobile devices and subsequently suffer from higher response times due to the wide-area network (WAN) \cite{Satyanarayanan09}.
In order to provide low latency response, the concept of Cloudlets were introduced. As shown in Figure \ref{fig:distributedCloud}, Cloudlets introduce additional compute resources near the mobile client to enable rapid instantiating of service software on the nearby compute resources. The Cloudlets provide a reliable uplink to the cloud as well as a wireless connection and computation resources to mobile clients. 

Virtualization techniques on the Cloudlets are leveraged to serve customized services for mobile clients on Cloudlets located in close proximity.
Satyanarayanan et al. \cite{Satyanarayanan09} also emphasize the capability to live migrate virtual machines between Cloudlets: Through this enabling strategy it is possible to make services accessible on a different compute device nearby the thin client.

\paragraph{Edge / Fog Cloud}
The provision of additional compute layers into the infrastructure, as defined by Cloudlets, can be further relaxed. Figure \ref{fig:distributedCloud} shows the properties of introducing further layers of compute resources, connected arbitrarily, as Edge/Fog Cloud. Edge and Fog Computing describes an infrastructure deployment in which cloud computing capabilities are provided at the edge of the network and on heterogeneous devices \cite{Hassan18}. The terms Edge and Fog Computing are often used interchangeably since they describe a similar distributed cloud scenario.

Bonomi et al. \cite{bonomi2012fog} differentiate the Fog cloud to be not exclusively located directly at the edge (only next the client device), but is rather extended to a larger area, functioning as fog infrastructure. 
Similarly, the differentiation between the term Fog and Edge gets increasingly used interchangeable. For example, Shi et al. \cite{shi2016edge} define Edge as "any computing and network resources along the path between data sources and cloud data centers". The term Edge is therefore also used for larger areas between the clients and the traditional Clouds. 

Motivated through the global increase in mobile and Internet-of-Things (IoT) traffic, Edge Computing is one of the key pillars of the upcoming 5G network standardization. To reduce the upstream bandwidth requirements of mobile devices and sensors, edge computing aims to execute computational tasks as close as possible to the data sources. This results on the one hand in reduced latencies for mobile devices and on the other hand prevents network congestion. Furthermore, location-aware applications are enabled and the infrastructure is more flexible because services can be deployed at several locations.

\subsubsection{Distributed Heterogeneous Cloud}
Connected things, such as IoT devices, are gaining in complexity and capacities in term of processing power. This further intensifies the need for these to go beyond rigid basic programming models to become fully networked computing systems capable of delivering advanced behaviours and interacting with their surroundings. The foreseen slow down in Moore’s law calls for an approach which progressively takes better advantage of available resources by exploiting heterogeneity and coping with the necessary balance among resources in high demand: computing and energy. 
In contrast to Cloudlets,
Edge and Fog Computing architectures already leverage
heterogeneous devices and can therefore be seen as first distributed heterogeneous Clouds. However, the availability of accelerators like GPUs, TPUs or FPGAs introduce further heterogeneous complexity to be considered in future environments.

\begin{figure}
    \centering
    \includegraphics[width=0.50\textwidth]{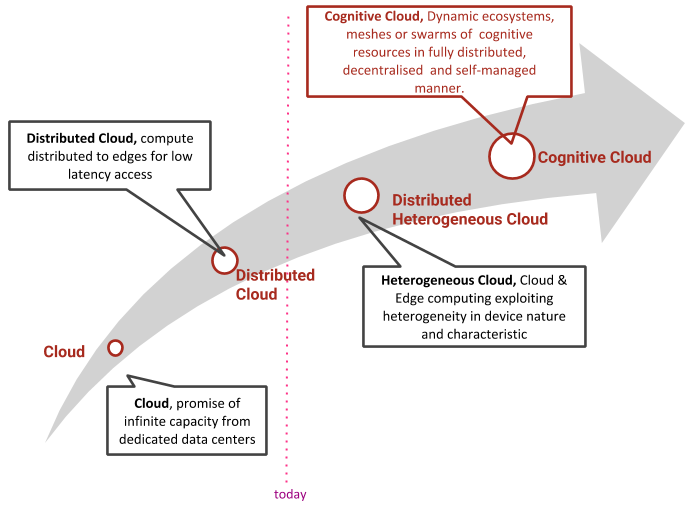}
    \caption{State of research and industry utilization of cloud concepts.}
    \label{fig:my_label2}
\end{figure}

We see traditional Clouds as well as distributed Cloud environments heavily applied in industry setting nowadays, while research is at the edge to discover the concepts of distributed heterogeneous Clouds as shown in Figure \ref{fig:my_label2}. Cognitive Clouds are stressing limitations of the preexisting concepts to ensure even more flexibility in highly dynamic use cases.

\section{Cognitive Cloud Characteristics and Principles}\label{principlesCCC}
Cognitive Compute Continuums aim to enhance the traditional Cloud environments by including a set of important characteristics and principles.

\subsection{Central Characteristics}
 
The central characteristics of a Cognitive Compute Continuum are remote sensing, ad-hoc capabilities, fully decentralized operation, and self-management layers.

\textit{Sensing:}
 Cognitive Compute Continuums make sense of and interact with the physical world. This environmental awareness and actionability is provided by sensors and actuators of edge devices. Thus, the cognitive edge of CCC enables cyber-physical systems, allowing for tasks to be tackled in the physical world and in edge devices' local surroundings. 
Beyond sensors and actuators, edge devices also provide cognitive capabilities that can be utilized to process sensor data streams, implementing for instance low-latency control loops at the edge. At the same time, the devices are connected to other nodes in their immediate surrounding and also clouds, allowing to dynamically activate more capabilities as needed.

\textit{Ad-hoc character:}
Cognitive compute continuums comprise of a variety of resources and are by nature dynamically changing. They consists of distributed and heterogeneous edge devices, edge and fog computing resources, as well as more centralized cloud resources. 
Resources closer to the edge are more distributed, more limited in their compute and storage capabilities, often also battery constrained, but at the same time closer to sensors and actuators, when compared to more centralized cloud resources. Distributed edge resources are also more dynamic. There is usually a high degree of mobility and node churn with edge resources that can attach and detach themselves to CCCs at any time. Edge devices might have to leave due to, for example, drained batteries, higher priority primary functions evicting CCC tasks, and failures. Cloud resources on the other hand are more stable, given the high levels of availability provided by today's data center technology, yet their availability depends on working communication links, where connectivity changes dynamically as well. 

\textit{Decentralized Operated Cloud:}
Cognitive Compute Continuums operate efficiently and reliably without a central entity for management and full knowledge about available resources and current workloads. Instead, nodes of the continuum work in a decentralized manner, utilizing local knowledge of and locally exchanging information on resources and workloads. Decentralized resource managers default to local resources as long as the current knowledge of resource availability and workload demands indicate feasibility under the given constraints and objectives of tasks, otherwise offloading tasks to connected resources. Local state and models of decentralized resource managers is continuously updated, yet information is expected to be incomplete and outdated. Therefore, CCCs make use of mechanisms to rectify situations, in which local information did not adequately reflect the current overall system state, requiring for instance rollbacks, re-submission, and re-scheduling. Such a decentralized and autonomous approach is necessary in light of the dynamic and ad-hoc nature of CCCs and their scale. Consequently, the resource management of CCCs is resilient against being temporarily or permanently partitioned into multiple continuums by network outages.

\textit{Self-management:}
Cognitive compute continuums apply autonomous computing techniques and machine learning to be able to adapt to unforeseen situations and anticipate future circumstances. That is, cognitive resources are self-aware, self-managed, self-protected, self-healing, and self-optimizing. These features are enabled by continuous monitoring and modeling of and by all nodes of the continuum. The nodes continuously update and apply models of current and future resource availabilities, node churn and mobility, system health and anomalies, as well as of workloads and task performance. The nodes of CCCs also continuously exchange information and models among each others for optimal decentralized decisions despite the incomplete global information and possibly outdated local knowledge, given the adhoc and dynamic nature of CCCs. This way CCCs do not need to rely on human operators, ultimately allowing for larger scales, quicker responses, and more efficiency.
    
\subsection{Key Principles} \label{sec:fwk}

Cognitive Clouds are based on three so-called AAA principles -- derived from IoT principles -- as an acronym for the Aware, Autonomous, and Actionable characteristics \cite{oriwoh2013guidelines, sun2016edgeiot}, defined as follows:

\subsubsection{Aware}
The sensing part of a Cognitive Cloud has to be able to perceive data or information about physical or digital nearby surroundings and has the ability to store this information to identify patterns and model changing circumstances.
Such dynamically changing environments are captured in the two models of resource awareness and context awareness. 

\paragraph{Context awareness}

\textit{Spatial Context awareness - Location:}
A key strategy, which is needed in a CCC is the foreseen availabilities of participatory devices in close spatial proximity.
This concept relies on the Edge computing attribute that considers physical proximity as the method that facilitates latency enhancement. While the consideration of physical location of the IoT devices and Edge nodes that constitute an Edge infrastructure are much of interest in certain usage scenarios. 
Key features to be implemented by a CCC are the determination of the physical areas in which a participatory devices of the CCC operates; the management of the CCC overlaps in physical locations; as well as physical node and network discovery mechanisms.

\textit{Social Context awareness - Resource Discovery and Collaboration:}
A participant (device) of a CCC is aware of further participants in close proximity. Some are function as allies aiming to collaborate and fulfill jointly tasks, while others pursue other tasks and competing against resources as adversaries, but might also spy on information, jam the transmission or block the system.
The collaboration can be defined in several ways. The traditional approach foresees a common master plan followed by different participants. Parallel or pipelined processing of a workflow are typical examples for a processing pattern within a swarm. Another option is to view the collaboration as a safety-network, where components can take over (parts of) the processing tasks due to failures or overload and thus mask the malfunctioning of the system. A further option is to use the other peers as a source of knowledge, for example what can a close participant tell me about its environment, to e.g. share valuable values or ML models to avoid coldstart situations. Finally, components with trust relationship to each other can jointly step against an adversary and develop on the fly a rescue plan or follow one pre-defined plan, which again leads to the very first point. The pre-defined tasks are associated with QoS KPIs, which participants try to optimize in a swarm.

\paragraph{Resource Awareness} Cognitive clouds need to be aware of the available resources to be able to efficiently provision tasks with the required compute, communication, and storage capabilities. Some resources available in the continuum of cognitive clouds are static and, aside of failures, permanently available. Other resources are dynamic. These resources can join and leave the cloud continuum at any time, with continuous churn due to user behaviour and node mobility. Resources that are not exclusively dedicated to cognitive clouds can furthermore dynamically adapt which capabilities and how much resources they provide, allowing to prioritize primary device functions with fluctuating resource demands. As the available resources can, therefore, change at any time, a full and up-to-date knowledge of the resources cannot be assumed. At the same time, knowledge of the available static and dynamic resources should be updated and exchanged continuously. That is, static resources can be captured using resource descriptions that denote the heterogeneity of resources, containing indicators for their performance for estimating task performance. Dynamic resources on the other hand require more continuous approaches to monitoring and performance modeling. %
		
\subsubsection{Autonomous} %
Resources in Cognitive Clouds and Cognitive Cloud infrastructure itself need to act as autonomous entities. They are responsible for their own configuration, monitoring, and management in a manner that exercises control over their own actions.
The autonomous characteristic of CCCs therefore apply to the resource management as described in the awareness section and further extends to scheduling approaches: 
Tasks in a CCC environment should be scheduled decentralized, without the dependence on a central entity.

Furthermore, the entities in a CCC need to be provided with autonomous mechanism in regards to fault tolerance: Due to distributed and ad-hoc nature, they demand a distributed monitoring approach which can be leveraged for self-healing purposes \cite{10.1007/978-3-642-10665-1_5}. Such characteristics of self-management were proposed by Paul Horn as Autonomic computing \cite{Horn2001AutonomicCI}. This also includes methods for self-configuration \cite{athreya2013designing} as well as self-protection \cite{wailly2012vespa} in terms of security and can thus be summarized as general autonomous self-optimisation \cite{nallur2009self}.

\subsubsection{Actionable} %
Actuation is not only dependent on the environmental context, but also involves the capacities of learning from past situations, the ability to predict and simulate as well as to enable real-time automation of actions. Such actions refer to tasks, which are carried out within the CCC. These typically include SLA optimization. Thus, CCC intelligently schedules and foresees need in e.g. infrastructure adaptability, data transmission, analytics adaptability to fulfil tasks. Additionally, CCC should be capable of foreseeing future workloads, so that the elasticity of services is ensured with respect to probable changes of KPIs.

\section{Cognitive Cloud Realisation}
\label{CCRealisation}
\begin{figure*}
    \centering
    \includegraphics[width=0.9\textwidth]{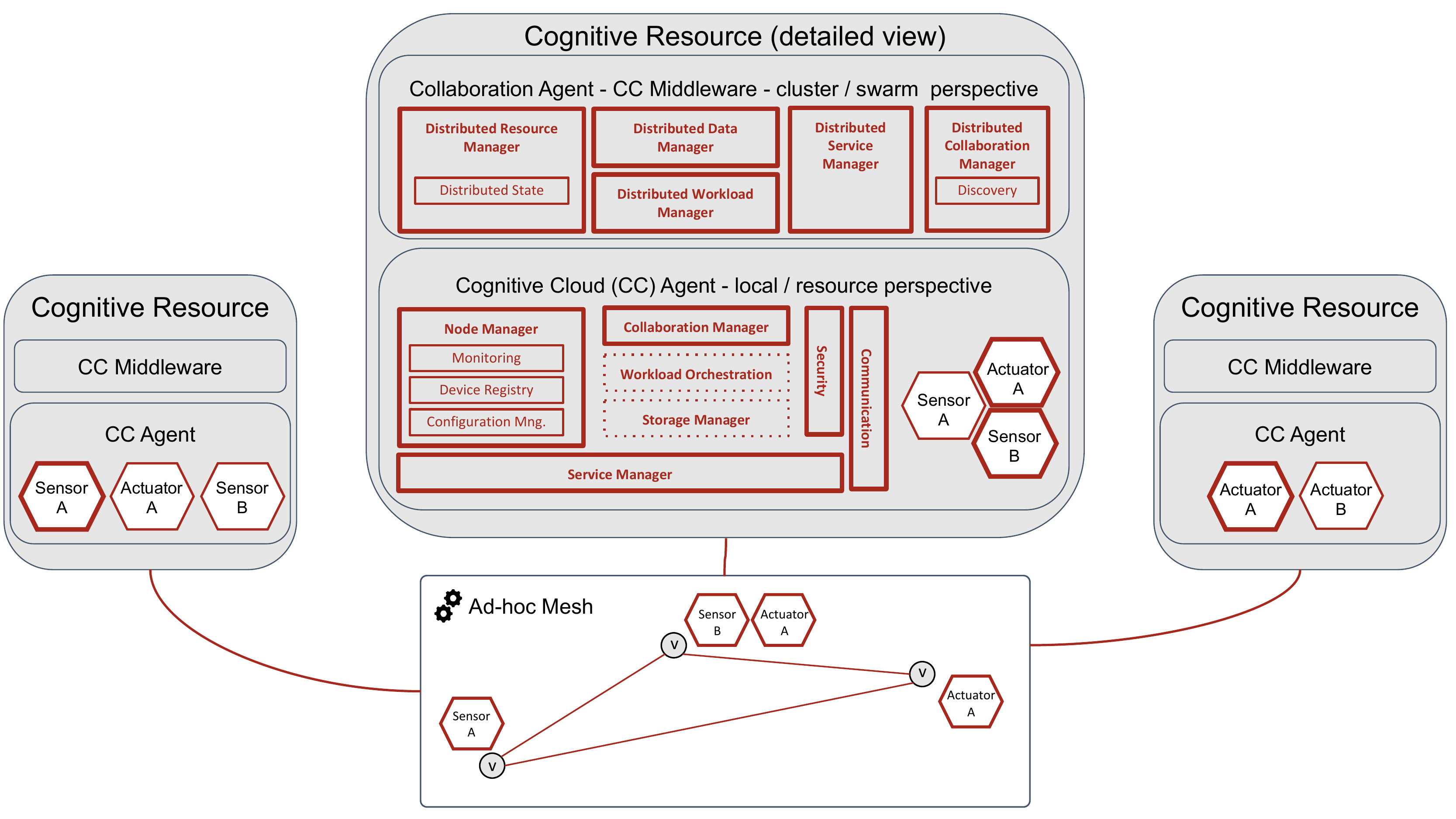}
    \caption{Realisation of the cognitive cloud.}
    \label{fig:my_label3}
\end{figure*}
The guiding idea of CCC is a Peer-2-Peer (P2P) system using an overlay network of autonomous entities for sensing, computing, and decision making based on ML for coordinated response to common tasks. CCC extends the request-reply paradigm of distributed heterogeneous cloud environments such as Edge and Fog environments, to a peer-to-peer paradigm.

Overall, CCCs are proposed as dynamic ecosystems, swarms, of cognitive resources operate in fully distributed, decentralised and self-managed manner. Swarms collectively aggregate cognitive compute resources and complement local knowledge through connection to other resources in the environment. Swarms encompass heterogeneous cognitive compute resources in IoT devices, edge resources and clouds.  In swarms, each of these resources add to the collective their specific capability and knowledge. Swarms are subject to certain degrees of resource availability instabilities, due to probability of unreliable resources’ network links and limitations in terms of battery and compute and storage capacity. 
 
In a swarm, devices automatically discover nodes in the proximity to form clusters of devices (small CCC clouds) in which they collaborate to fulfil tasks. They coordinate themselves in a fully decentralised manner, without a single source of information and command. Nodes can join and leave the swarm at any time and swarms can be merged or divided when necessary for a given task. 

The architecture of a Cognitive Cloud instance, a Swarm,  is depicted in Figure \ref{fig:my_label3}. It is organised using two main building blocks: 
\paragraph{Cognitive Cloud(CC) Agent} manages the resources at the level of a local participant cognitive device. It offers the capabilities which enables the resource to participate in a swarm and enables perception of the environment by means of data gathered from device’s sensors. 
\paragraph{Collaboration Agent} builds the swarm perspective and offers features to discover nodes, the overall coordination among CC Agents for the realisation of the Cognitive Cloud middleware and builds the distributed registries among participant resources that constitute the P2P network. 

Both building blocks are deployed at each of the participant IoT and Edge devices and Cloud infrastructures building a P2P network by means of the Collaboration Agent features among them. 

\subsection{Cognitive Cloud (CC) Agent}
The CC Agent building block is constituted by the following components.
\paragraph{Node Manager}
Node Manager aggregates all capabilities to handle a cognitive resource that takes part in the \textit{swarm}, enabling the node to be managed as part of it. Node Manager offers an unified abstract interface that allows to  handle all resources in a similar manner despite of their specific heterogeneous characteristics. The Node Manager's Configuration manager allows bootstrapping of the resource in the infrastructure by defining its suitable parameters. It can consider aspects such as position, available resources and communication gateway. The Device registry permits to handle the associated sensors to the device in order to make generated data accessible to the Storage manager and to enable context awareness. Associated sensors can be external or embedded into the device.
Monitoring enables to have a complete view on the status of the resource and its associated sensors. Building on top of gathered monitoring information, diverse profiles for the resource can be constructed to be consumed by the rest of components in the CC architecture: profiles about its dynamic status (utilisation rate, available battery, tasks schedule, physical position);  profiles containing its Static hardware characteristics i.e. considering the type of CPU, available memory and network links; or a profile relying on Static software (OS, middleware, ML algorithms and protocols).

\paragraph{Service Manager}
This component controls the status of all services and tasks in execution in a certain device. By means of the Node Manager monitoring information, it oversees the consumed resources of tasks execution in the host device and can take corrective actions i.e. in terms of ensuring the QoS and Fault tolerance at local level, or in coordination with Collaboration manager, at overall infrastructure levels. This behaviour is coordinated by the self-management policies described for the CCC infrastructure.  

\paragraph{Collaboration Manager}
This component offers the necessary mechanisms to discover available resources in a location and to publish the cognitive resource features made available to the rest of the infrastructure. Each resource in the infrastructure details by means of a predefined syntax the typologies of tasks it is able to execute, as well as, the data sources that can be obtained of it. The Collaboration Manager component enables self-awareness for the Cognitive Cloud, by modelling the information sharing among swarm members. It is also a key element to offer self-organisation to the degree it defines the way of interaction among a resource and the rest of the CCC by the Collaboration Manager at the Collaboration agent level.

\paragraph{Workload Orchestrator}
This component is responsible for managing the lifecycle of the tasks and applications to be executed by the cognitive resource. It relies on existing baseline to deploy services in agents enabled with  Docker, Docker Swarm, or Kubernetes. Tasks deployed in Docker engines can be conventional docker images, docker-compose, or pod's files depending on the technology choice made the infrastructure. Status of the execution of tasks are overseen by the Service manager at the resource level. 

\paragraph{Storage Manager}
The Storage Manager controls the processes associated to the sharing of data collected by the cognitive resource and associated sensors and the rest of the swarm by means of Collaborative Agent Distributed Data manager. It can be implemented as a distributed data index making use of distributed storage systems such as Etcd\footnote{\url{https:/etcd.io/}} or Apache Cassandra\footnote{\url{https:/cassandra.apache.org/}}. It can also define specific data synchronisation processes among the participant cognitive resources in the swarm. In certain scenarios, it will be key to handle  appropriately diverse levels of data access and encryption by interaction with security components. 

\paragraph{Security}
Security is considered to be a component which is very much dependent in the specific use case in which this architecture will be applied. Diverse levels of security requirements and resource self-protection mechanisms can be considered depending on the specific needs ranging from basic access control to full encryption of data and communications.

\paragraph{Communication}
Similarly, the heterogeneity of resources that can take part on a swarm bring the need of considering a diversity of communication protocols which will be defined in the specific usage scenario and devices in the specific instance. In addition, mesh networks are an essential enabler for swarms which collaboratively with resource discovery features at Collaboration manager have to be able to establish swarms. \color{black}

\subsection{Collaborative Agent}
The Collaborative Agent building block enables the orchestration of resources managed by CC Agents in order to form and act coordinately as a swarm of compute resources. The main components envisaged in this building block are described below. 

\paragraph{Distributed Resource Manager}
The Distributed Resource Manager Component allows the generation of swarms out of the available set of CC resources. In this context the Distributed State allows to have a collective view of all CC Agents available at a certain point of time in the CCC infrastructure. The Distributed Resource Manager relies in the Distributed Collaboration Manager to discover the Nodes that form the swarm. 

By building on a distributed storage among all CC Agents Node Manager's information it offers a synchronised view among nodes. This view is kept up to date via the inherent consensus algorithms in distributed storages, for instance based on Raft algorithm and Etcd. The fact of using consensus algorithms also supports the CCC need for managing dynamic resource availability. CCC target resources are by definition not dedicated to the infrastructure and therefore, they are prone to loose connectivity, for instance due to battery deplete or use of unstable network links. 
Overall, the Distributed Resource Manager Component allows self-configuration of swarm behaviour (controls, scheduling, access, fault tolerance and service life-cycle management) over services and tasks to be executed in the swarm. 

\paragraph{Distributed Service Manager}
It enables allocating services available to the most suitable resources producing optimal performance and efficient use of resources. The flow of actions within this component is the following: First, this component identifies these services in the most suitable resources available (discovered via  Distributed Collaboration manager) and it then manages the lifecycle of services execution and termination of them by interacting with Distributed Workload Manager. Finally, it also manages the QoS of the service being executed, being able to perform policies defied self-management actions in case the expected QoS is not achieved or a failure is detected in an execution resource at CC Agent Service Management. Overall, the Distributed Service Manager aims at building an autonomic and  resilient service provision relying on resources and services profiling and forecasting.

\paragraph{Distributed Workload Manager}
The Distributed Workload Manager is responsible for deploying and managing the services in CCC infrastructures which can be composed of multiple resources in collaboration with Workload Orchestration components at CC Agent level. 
Tasks and services can be packaged among others as containers or software functions.

\paragraph{Distributed Collaboration Manager}
The Distributed Collaboration Manager is able to select from the all available resources, those that can execute the service "more optimally" based on a set of established collaboration patterns and considering the typologies of tasks each kind of resource is able to fulfil (according to the information published by each CC Agent Collaboration manager). In order to perform the selection of nodes multiple criteria can be utilised: depending on resource foreseen availability; based on the overall predicted QoS for the service and task execution for each resource; as well as, looking for optimising locality of service and tasks execution. All these optimisation criteria heavily depend on CC Agent Node Manager resource profiles. 

\paragraph{Distributed Data Manager}
The Distributed Data Manager aims at optimising data access by tasks and services. Overall, it has the objective to provide an abstraction layer which permits services to use available data sources ( declared by CC Agents Storage Services), handling data replication and movements transparently and optimally in order to enhance tasks execution performance.

\section{Related Work} 
The utilisation of cognitive and machine learning techniques for resource and overall infrastructure management across Edge and Cloud continuum has recently gained significant attention in the research community. Le Luc in \cite{Duc2019} explores the application of these techniques for reliable resource provisioning for edge-cloud applications, focusing on workload analysis and prediction, placement and consolidation, elasticity and remediation. It performs a literature review of works in the area of application placement at the level of existing resources in data centre installations in Cloud environments and mobile Cloud computing computation offloading. The latter typically refers to determining the parts of an application on a mobile device to be transferred to more powerful Cloud environments,  while the approach relies on devices in the vicinity for a dynamic and opportunistic compute environment, as envisioned by us.

More in detail, Ilager in \cite{Ilager} proposes a conceptual model for AI-centric resource management systems in distributed systems that encompass Edge and Cloud. It explores two scenarios in which to apply AI techniques: a resource management of large scale data centre for energy optimisation and Microsoft Azure Cloud; and a AI based mechanism for configuration of GPU Clouds for workload scheduling. These scenarios show the potential of the employment of AI for resource management in Clouds, however it misses the complexity and dynamicity that the consideration of Edge computing resources inevitably adds. Rodrigues in \cite{ Rodrigues2020} explores works related to the application of AI to Mobile Edge computing, advocating for the use of machine learning over heuristics and convex optimisations, yet recognising the need for more research to enable Edge device collaboration, as we foresee in our work. 

At the level of sensing, \cite{Wang2019} develops the concept of Social Edge Intelligence, that proposes the integration of artificial intelligence with human intelligence to address critical research challenges of Edge computing. In this context, it proposes the challenge of efficient resource management that exploit the heterogeneity present in the Edge devices and diagnoses the need of additional research to enable seamless device collaboration for timely task execution.   

The concept of Edge Clustering has been recently studied by \cite{Babou}, presenting a Hierarchical Load Balancing and Clustering Technique for Home Edge Computing. The proposed approach considers a three-layer architecture (Cloud, MEC, Home Edge Server). If a resource gets overloaded, it forwards this request to the next layer in the stack. This approach was also taken in mF2C project \cite{mf2c}, considering a two-level Hierarchy (Fog, Cloud). In contrast, CCCs do not assume any pre-defined layered structure among resources, solely basing placement decisions on expected nodes availabilities and  performance. 

\section{Conclusion}

In this paper we presented our vision towards the realisation of Cognitive Computing Continuums and introduced an architecture for implementing this vision. Our idea of CCCs aims at bridging the existing gap between Edge and Cloud resources, fully exploiting the computing capacity available in the heterogeneous cognitive devices at the Edge and facilitating a seamless collaboration with dedicated Edge and Cloud resources. Our vision is complemented with an architecture for implementing this goal. In the future, we will concentrate on investigating resource management and scheduling when the sensing abilities of CCCs are used for AI.

\bibliographystyle{IEEEtran}
\bibliography{publications}

\end{document}